\documentclass[prd,onecolumn,amsmath,amssymb,floatfix,nofootinbib]{revtex4}

\usepackage{amssymb}
\usepackage{amsmath}

\usepackage{graphicx}
\usepackage{graphics}
\usepackage{dcolumn}
\usepackage{color}
\usepackage{rotate}
\usepackage{fancyhdr}
\usepackage{hyperref}
\usepackage{indentfirst}

\usepackage{umoline}
\usepackage{ulem}

\def\be{\begin{eqnarray}}
\def\ee{\end{eqnarray}}
\def\ba{\begin{eqnarray}}
\def\ea{\end{eqnarray}}

\definecolor{darkred}{rgb}{.743,0,0}

\begin{document}
\title{Cosmic microwave background constraints on primordial black hole dark matter}

\author{Daniel Aloni$^{1}$}
\author{Kfir Blum$^{1}$}
\author{Raphael Flauger$^{2}$\vspace{0.2cm}}
\affiliation{$^1$Weizmann Institute, Department of Physics and Astrophysics, Rehovot, Israel 76100\\
%\affiliation{
$^2$University of California, San Diego, 9500 Gilman Drive 0319
CA La Jolla 92093}
%\date{\today}

\vspace{10 cm}
\begin{abstract}
We revisit cosmic microwave background (CMB) constraints on primordial black hole dark matter. Spectral distortion limits from COBE/FIRAS do not impose a relevant constraint. Planck CMB anisotropy power spectra imply that primordial black holes with $m_{BH}\gtrsim 5~M_{\odot}$ are disfavored. However, this is susceptible to sizeable uncertainties due to the treatment of the black hole accretion process. These constraints are weaker than those quoted in earlier literature for the same observables. 
\end{abstract}
%\pacs{}

\maketitle
%\tableofcontents
%%%%%%%%%%%%%%%%%%%
\section{Introduction and result}
Primordial black holes (BHs) accrete matter in the early Universe, releasing accretion luminosity that heats and ionises hydrogen leading to potentially observable effects in the spectrum and anisotropy pattern of cosmic microwave background (CMB) radiation. 
Ref.~\cite{Ricotti:2007au} analyzed these effects, concluding that BHs with mass $m_{BH}\gtrsim5\times10^{-2}~M_{\odot}$ are excluded as dark matter (DM) candidates (see e.g.~\cite{Carr:2016drx} for a recent review). 
We revisit cosmological aspects of the analysis, finding weaker limits. Considering spectral distortions we find no constraint\footnote{This is consistent with footnote in Ref.~\cite{Clesse:2016ajp}.}.
Considering CMB anisotropies, using Planck data we find, for BHs making up all of the DM, likelihood ratios of $10^{-1}$ and $10^{-2}$ for $m_{BH}=4.9~M_{\odot}$ and $m_{BH}=6.1~M_{\odot}$, respectively. For reference, for a Gaussian likelihood a likelihood ratios of $0.14$ and $0.011$ correspond to $2\sigma$ and $3\sigma$ constraints, respectively.

%
%\be\label{eq:1} m_{BH}<5~M_{\odot},\;\;\;{\rm 95\%CL.}\ee
%
The limit on $m_{BH}$ is susceptible to theoretical uncertainties that are difficult to quantify, at the level of a factor of few at least, due to the modelling of BH accretion rate and luminosity. 

A recent re-evaluation of the CMB anisotropy constraint was presented in Ref.~\cite{Chen:2016pud} which, however, repeated the accretion analysis of~\cite{Ricotti:2007au} including inaccuracies that we explain below, especially regarding the relative bulk velocity between the BH and the baryonic plasma. 

While this paper was being typed for publication, Ref.~\cite{Ali-Haimoud:2016mbv} appeared, dealing with the same topic. Compared with our simple analysis that adopts the accretion modelling from Ref.~\cite{Ricotti:2007au} and refines the cosmology, Ref.~\cite{Ali-Haimoud:2016mbv} took on the more challenging task of also redoing the accretion astrophysics. The bound on $m_{BH}$ found in that work is weaker than ours by a factor of between 2 to 20, depending on the details assumed in the description of the accretion process.

%%%%%%%%%%%%%%%%%%%
\section{Analysis}\label{sec:1}

We modify RECFAST~\cite{Seager:1999km,Wong:2007ym} to include the BH accretion luminosity in the cosmological ionization history, and use CAMB~\cite{Lewis:1999bs,Howlett:2012mh,camb_notes} to calculate the effect on the CMB anisotropies. 
We use the formulae in Ref.~\cite{Ricotti:2007au} to relate the BH mass accretion rate to the sound speed and to the relative velocity between the BH and the plasma, and to parametrise the accretion luminosity per BH, $L_{BH}$, giving power emitted per unit volume
\be Q_{pBH}&=&\frac{\rho_{BH}}{m_{BH}}L_{BH}
\ee 
where $\rho_{BH}$ is the BH mass density.

\begin{figure}[htbp]
\begin{center}
\includegraphics[width=0.65\textwidth]{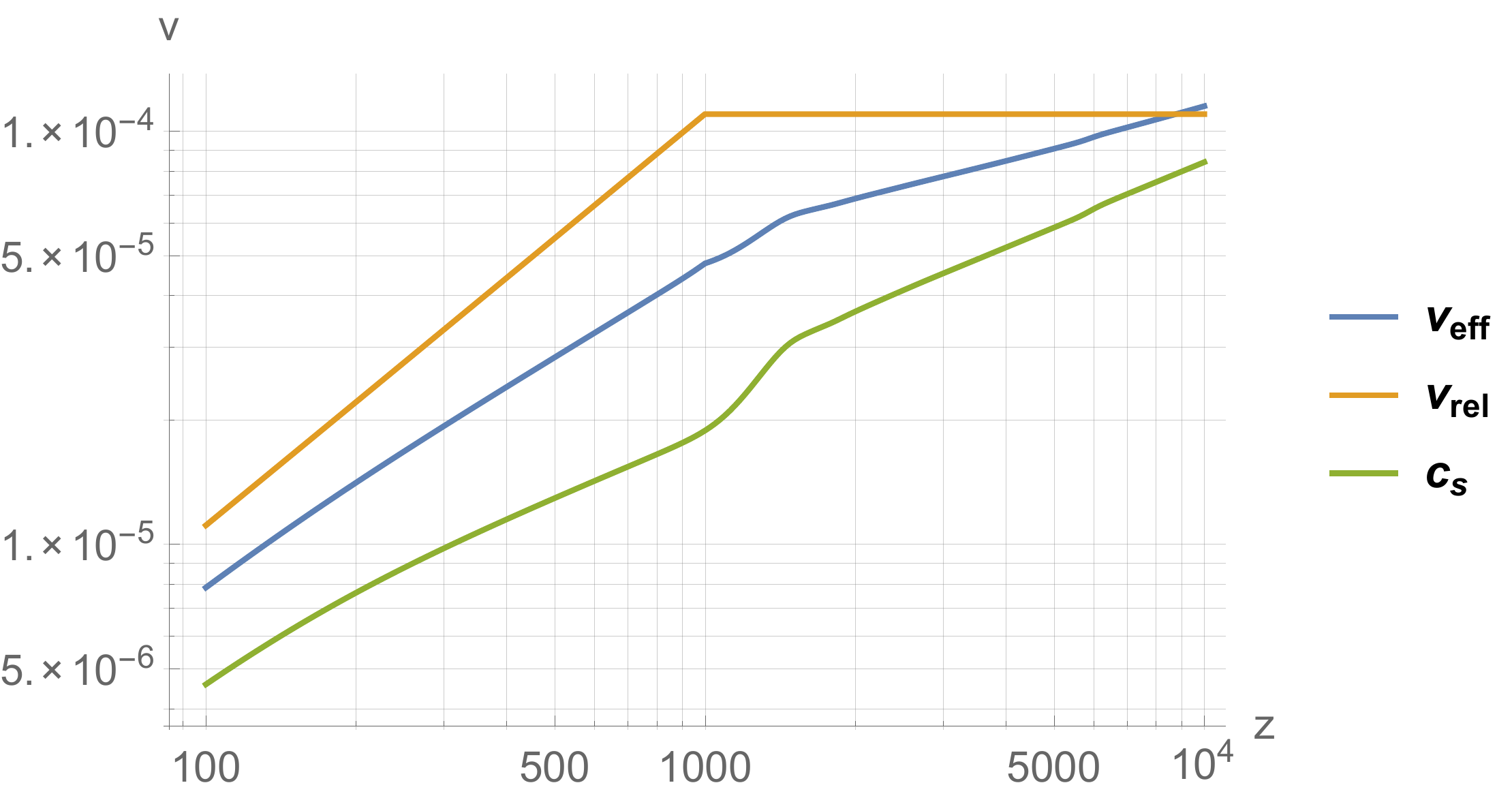}
\caption{Effective relative velocity between DM and baryons, in units of $c$. Blue: statistical average weighted by luminosity (as used in the analysis). Orange: RMS $v_{\rm rel}$. Red: speed of sound.}
\label{fig:Vrel}
\end{center}
\end{figure}

The relative bulk velocity $v_{\rm rel}$ between a BH and the plasma affects the BH accretion rate. Ref.~\cite{Ricotti:2007au} used the ``cosmic Mach number" defined in~\cite{1990ApJ...348..378O} to derive $v_{\rm rel}$. However, in the linear regime, we obtain a different expression for the RMS relative velocity that is given by~\cite{Tseliakhovich:2010bj}
\be\label{eq:linv}
 v_{\rm rel}^2=\int\frac{dk}{k}\Delta_{\xi}^2(k)\left(\frac{\theta_b(k)-\theta_c(k)}{k}\right)^2.
\ee
Here $k$ is the comoving wavenumber, $\Delta_\xi^2(k)\approx2.4\cdot10^{-9}$ is the input curvature perturbation variance per $\log k$, and $\theta_{b,c}(k)$ is the velocity divergence of the baryon and dark matter fluids~\cite{Ma:1995ey}. 
Eq.~(\ref{eq:linv}) was evaluated in Ref.~\cite{Dvorkin:2013cea} using CAMB. A simple analytic approximation of the result, that we use in numerical calculations below, is shown by the orange line in Fig.~\ref{fig:Vrel}. Comparing to Fig.~2 of~\cite{Ricotti:2007au}, our result for $v_{\rm rel}$ is larger by about a factor of five at $z\sim10^3$, leading to suppressed accretion. 

Following~\cite{Ricotti:2007au}, we define the Bondi-Hoyle effective velocity $v_{\rm eff}$ such as to incorporate the statistical nature of $v_{\rm rel}$ in cosmological perturbation theory and its interplay with the parametric dependence of the BH accretion luminosity\footnote{See Eq.~(13) in~\cite{Ricotti:2007au}.},
\be\label{eq:veff} v_{\rm eff}^{-6}&=&\left\langle \left(v_{\rm rel}^2+c_s^2\right)^{-3}\right\rangle,\ee
where $\langle.\rangle$ denotes averaging with a Maxwellian velocity distribution with RMS speed given by $v_{\rm rel}$.  
The resulting $v_{\rm eff}$ is shown by the blue line in Fig.~\ref{fig:Vrel}.

%%%%%%%%%%%%%%%%%%%
%%%%%%%%%%%%%%%%%%%

We are now in position to calculate the effect of BH accretion luminosity on the state of the plasma. Before reporting the results, we note that Ref.~\cite{Ricotti:2007au} used numerical simulations to estimate a number of effects involved in the calculation, that we treat more simplistically. The effects and our differences in treating them are as follows. 

{\bf (i)} BH accretion luminosity produces a spectrum of radiation. Photons of different energy induce different effects in the plasma, including ionisation, atomic excitations, heating, and the delayed subsequent absorption of redshifted X-ray photons. In place of the simulations of~\cite{Ricotti:2007au}, we simply assume that a fraction $(1-x_e)/3$ of the total luminosity $L_{BH}$ goes into instantaneous ionisation of the plasma, a fraction $(1-x_e)/3$ goes to atomic excitations, and a fraction $(1+2x_e)/3$ goes to heat. Assuming instantaneous deposition of the total BH luminosity in the plasma should lead to an over-estimate of the significance of the CMB constraint we derive, because (a) a large fraction of the photons emitted at $z<10^3$ escape to redshift $z=0$ without incurring any ionisations, and (b) the differential contribution to the Thomson optical depth $\tau=\int dz\delta\tau(z)$ scales as $\delta\tau(z)\propto(1+z)^{\frac{1}{2}}\delta x_e(z)$, so a given increase $\delta x_e$ in the ionised fraction contributes more to the optical depth at high $z$. 
%
%The choice of the partition of the luminosity deposited in each channel: ionisation, excitation, heating, also affects the bounds at the O(1) level.

{\bf (ii)} Ref.~\cite{Ricotti:2007au} included in their simulations back-reaction processes where heating of the plasma near the BH temporarily halts the accretion, leading to suppressed duty cycle of the emission. We neglect this effect, which could relax our bounds further.

{\bf (iii)} Ref.~\cite{Ricotti:2007au} considered cases in which BHs make up only a small fraction of the DM, $f_{BH}<1$ with $f_{BH}=\rho_{BH}/\rho_{DM}$. In this case, a dark halo of DM accreting into the BH increases the effective BH mass, leading to enhanced accretion luminosity. We neglect this effect. The cost is that one cannot trust our limits when $f_{BH}<1$. However, at the point $f_{BH}=1$ there is no room for a dark halo and our calculation determines the value of $m_{BH}$ below which BHs could make up all of the DM consistent with CMB data.

%{\bf (iv)} Ref.~\cite{Ricotti:2007au} modified the Bondi-Hoyle effective velocity of Eq.~(\ref{eq:veff}) to incorporate the statistical nature of $v_{\rm eff}$ in cosmological perturbation theory and its interplay with the parametric dependence of the BH accretion luminosity on $v_{\rm eff}$\footnote{See Eq.~(13) in~\cite{Ricotti:2007au}.}. We neglect this refinement, expecting that underlying uncertainties in the accretion and the modelling of the accretion luminosity are at least equally significant. Instead we use the formulae of Ref.~\cite{Ricotti:2007au} for the BH accretion rate with $v_{\rm eff}$ given simply by Eq.~(\ref{eq:veff}). For the Mach numbers of interest, of order a few, the difference in $v_{\rm eff}$ between our Eq.~(\ref{eq:veff}) and the implementation of~\cite{Ricotti:2007au} is at the level of 30\% to 70\% for the post-recombination epoch, relevant for CMB anisotropies, and smaller than that for redshift $z>10^3$ relevant for spectral distortions.

\section{Spectral distortions and anisotropy correlation functions}\label{sec:spec}
\begin{figure}[htbp]
\begin{center}
\includegraphics[width=0.65\textwidth]{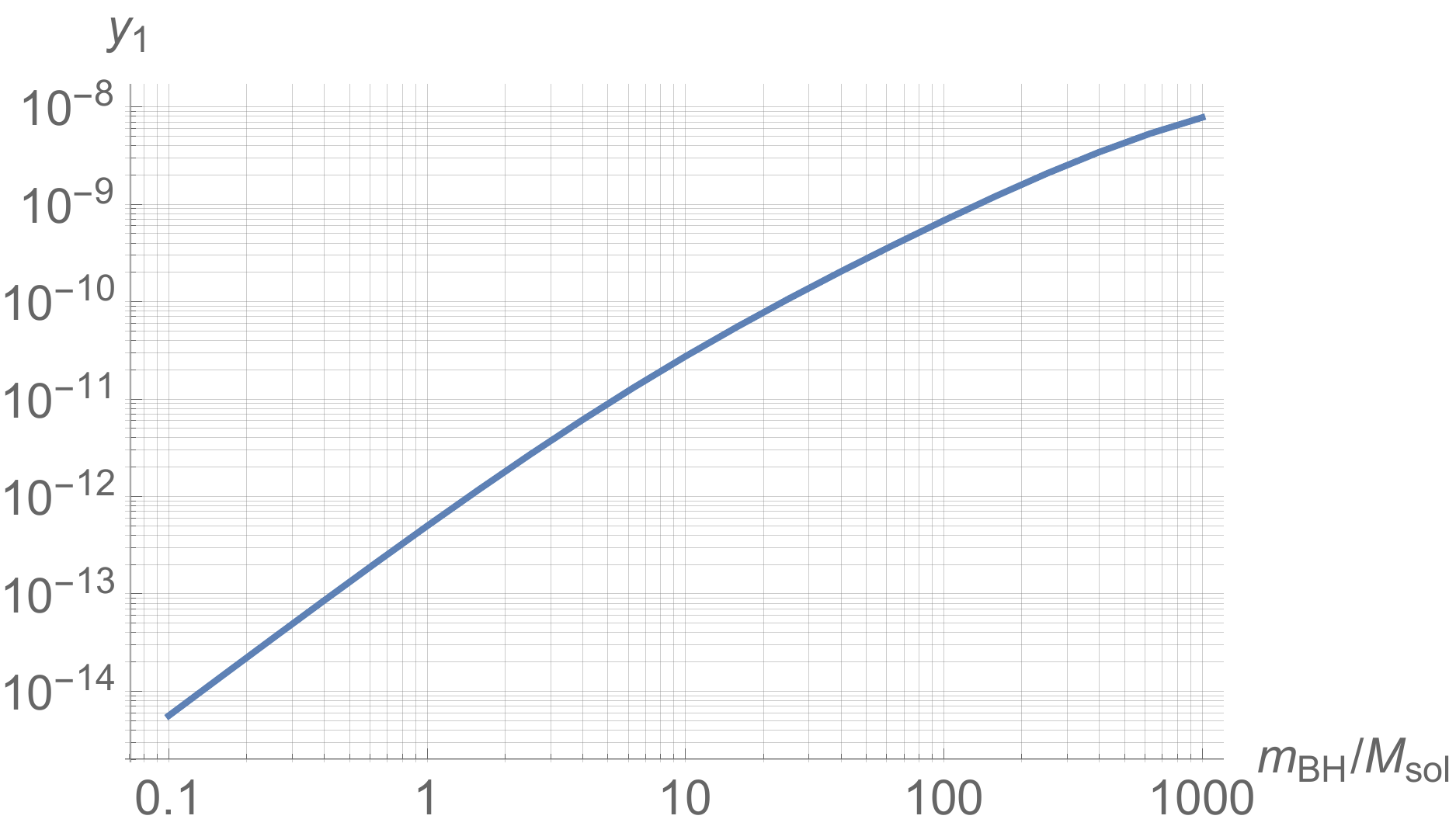} 
\caption{Spectral distortion $y$ parameter.}
\label{fig:y}
\end{center}
\end{figure}

%%%%
Ref.~\cite{Ricotti:2007au} calculated the contribution of BH accretion luminosity to the Compton $y$ parameter, that was estimated by the contribution $y_1$ obtained from energy injected in the redshift interval $z_{rec}<z<z_{eq}$. We can write this estimate as 
\be\label{eq:y}
y&\approx&0.25\int_{t_{eq}}^{t_{rec}}dt\frac{Q_{pBH}(t)}{U_{CMB}(t)}
=0.25\int_{z_{rec}}^{z_{eq}}\frac{dz}{H(z)(1+z)}\frac{Q_{pBH}(z)}{U_{CMB}(z)}.\ee
The result is shown in Fig.~\ref{fig:y}. The constraint from COBE/FIRAS~\cite{Fixsen:2009ug,CMBtemperatureCOBE} is $y<1.5\times10^{-5}$ at 95\%CL; the BH contribution cannot be constrained in the mass range of interest. %This conclusion persists also if we extend the integration of $y$ to the interval $10^2<z<10^5$, in which case we find that FIRAS excludes BH DM only for $m_{BH}>350~M_{\odot}$.
%%%%%%%%%%%%%%%%%%%
%%%%%%%%%%%%%%%%%%%

\begin{figure}[htbp]
\begin{center}
\includegraphics[width=0.475\textwidth]{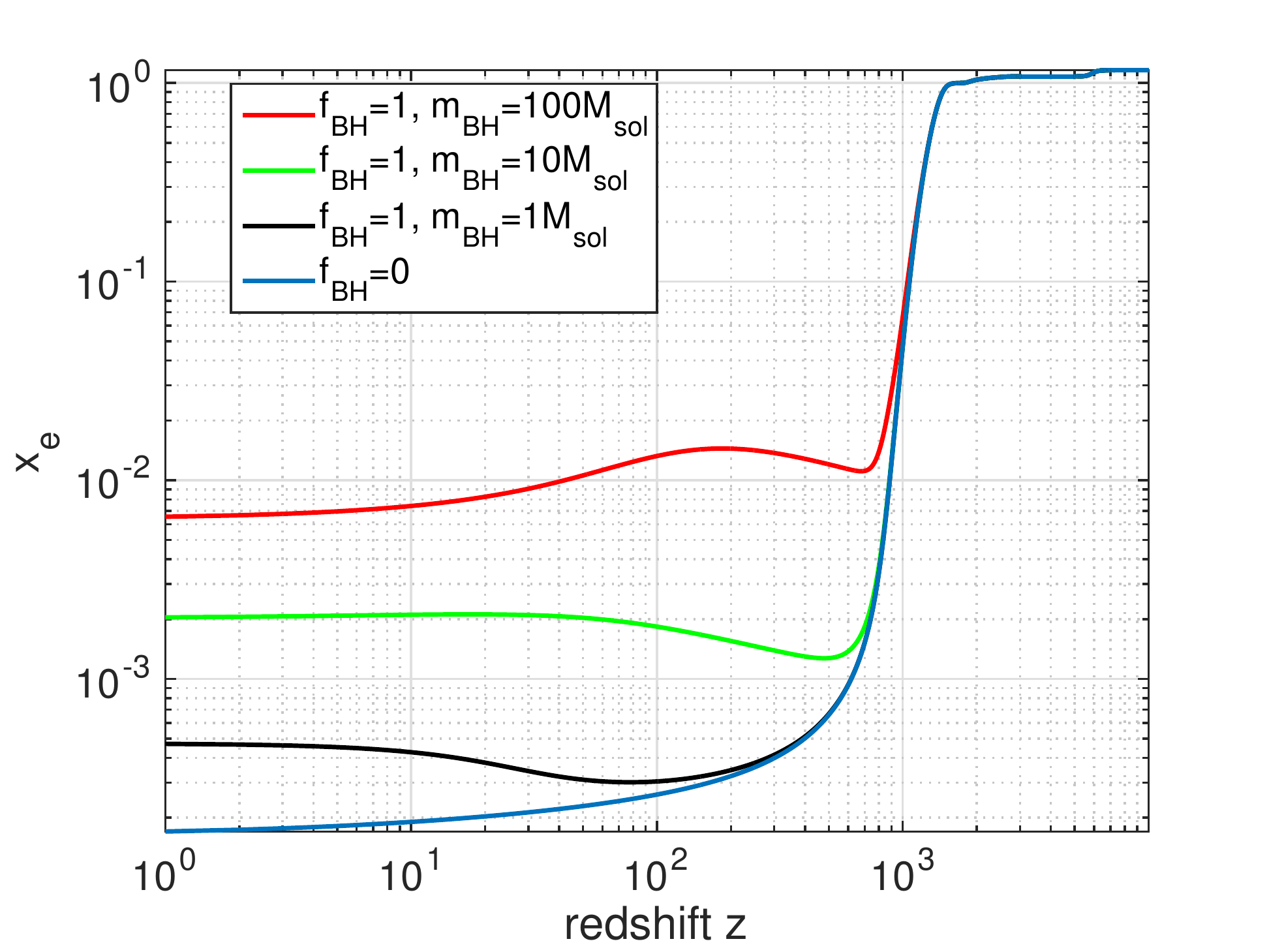} \quad
\includegraphics[width=0.475\textwidth]{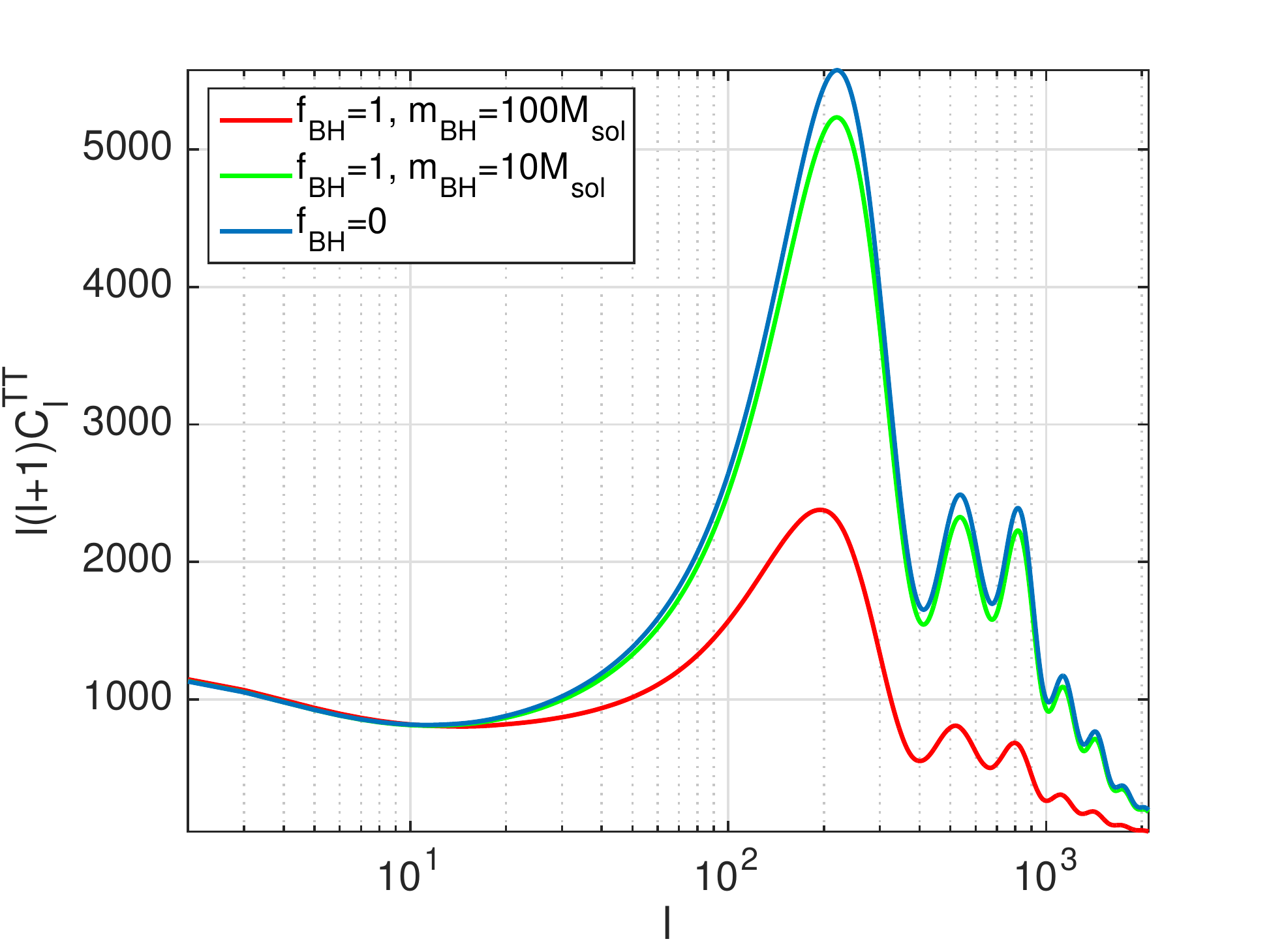}
\caption{Left: ionisation fraction. For clarity, the late re-ionisation at $z\sim$few is not shown here although it is included in the calculation of the CMB anisotropies. Right: TT power spectrum. The effect is difficult to see for $m_{BH}=10~M_{\odot}$.}
\label{fig:Cl}
\end{center}
\end{figure}

\begin{figure}[htbp]
\begin{center}
\includegraphics[width=0.9\textwidth]{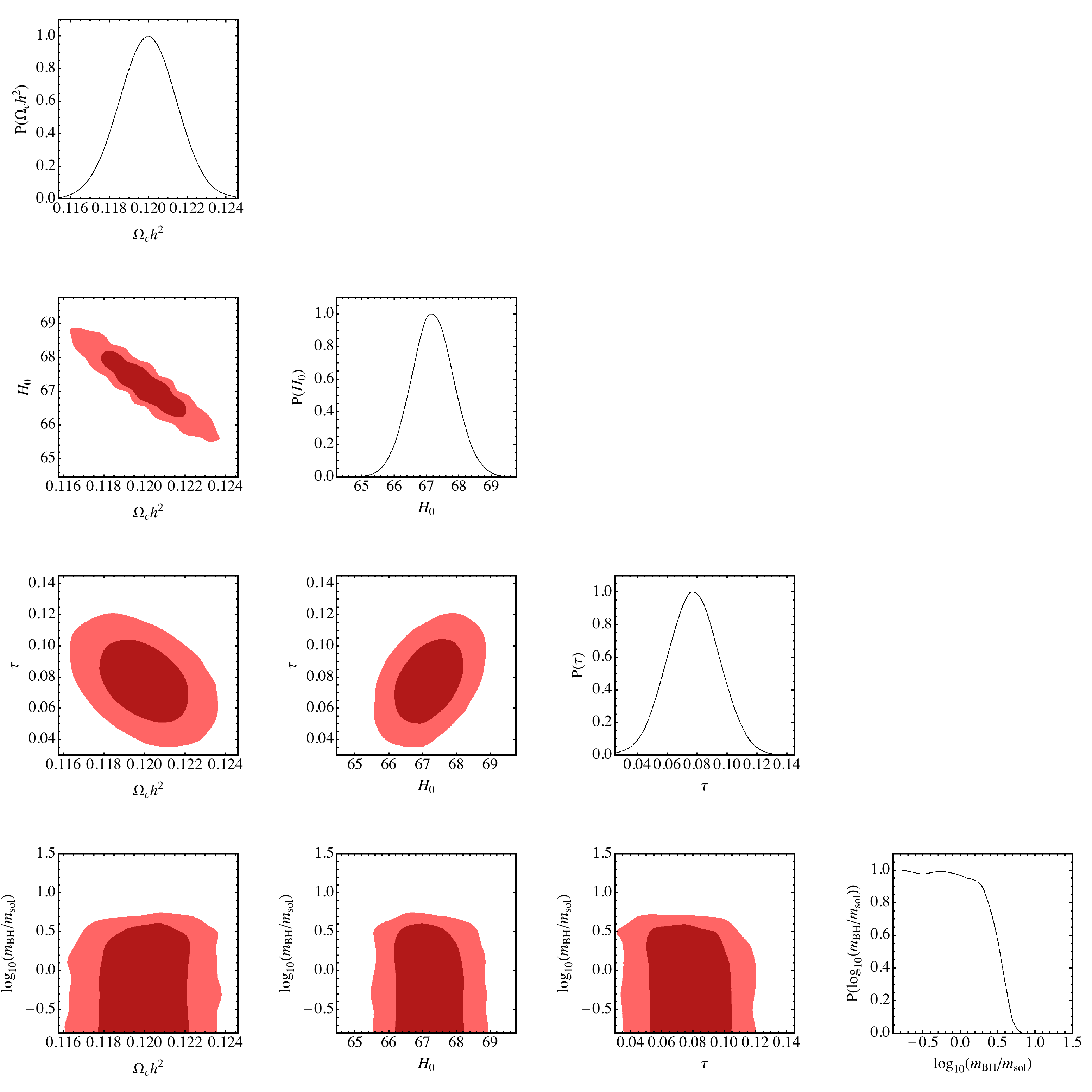} 
\caption{Likelihood plots for $\Lambda$CDM+BH model, using Planck TT, TE, EE data~\cite{Aghanim:2015wva,Ade:2015xua}.}
\label{fig:lik}
\end{center}
\end{figure}

% 
Moving on to CMB anisotropies, in Fig.~\ref{fig:Cl} we show results for the recombination history (left panel) and TT power spectrum (right panel) for sample values of $m_{BH}$. Performing a likelihood analysis~\cite{Dunkley:2004sv,cosmomc_notes} for the 6 usual $\Lambda$CDM parameters augmented by another parameter for $m_{BH}$, using the latest TT, TE, EE anisotropy data from Planck~\cite{Aghanim:2015wva,Ade:2015xua}, leads to the constraint quoted in the introduction. In Fig.~\ref{fig:lik} we show part of the likelihood triangle. 
%The BH effect is difficult to discern by eye in the TT power spectrum even for $m_{BH}=10~M_\odot$ (green curve, right panel).
%%%%%%%%%%%%%%%%%%%
%%%%%%%%%%%%%%%%%%%
%%%%%%%%%%%%%%%%%%
%%%%%%%%%%%%%%%%%%

%%%%%%%%%%%%%%%%%%%
\section{Summary}
We have re-analyzed the CMB constraints on primordial black holes (BHs) playing the role of cosmological dark matter. We find that primordial black holes with masses $m_{BH}>5$~$M_\odot$ are disfavored. This limit is subject to large, and difficult to quantify, theory uncertainty arising from the treatment of accretion and accretion luminosity of the BHs. Assuming, for concreteness, the same accretion prescription as in the earlier analysis of Ref.~\cite{Ricotti:2007au}, our limit is weaker despite the fact that we use Planck CMB data of far superior quality compared to the WMAP3 data considered in~\cite{Ricotti:2007au}. 

\vspace{6 pt}

\acknowledgments
We thank Jens Chluba and, especially, Yacine Ali-Ha$\ddot{\i}$moud for useful discussions. 
KB is incumbent of the Dewey David Stone and Harry Levine career development chair, and is supported by grant 1507/16 from the Israel Science Foundation and by a grant from the Israeli Centres Of Excellence (ICORE) program. RF is supported in part by the Alfred P. Sloan Foundation.

\bibliography{ref}

\end{document}